\documentclass[12pt]{article}
\usepackage{graphicx}
\newcommand{\beq}{\begin{equation}}
\newcommand{\eeq}{\end{equation}}
\newcommand{\bea}{\begin{eqnarray}}
\newcommand{\eea}{\end{eqnarray}}

\newcommand{\indi}{\nu_1\cdots\nu_{p+1}}

\newcommand{\indii}{\nu_1\cdots\nu_{p+2}}
\newcommand{\ha}{\hat{a}}
\newcommand{\hg}{\hat{g}}
\renewcommand{\b}{\beta}
\renewcommand{\a}{\alpha}

\newcommand{\ra}{\right\rangle}
\newcommand{\la}{\left\langle}
\begin{document}
\begin{titlepage}
\begin{flushleft}
       \hfill                      {\tt hep-th/000xxxx}\\
       \hfill                       FIT HE 4-01/00 \\
\end{flushleft}
\vspace*{3mm}
\begin{center}
{\LARGE Nonlinear sigma model and Yang-Mills theory\\ }
\vspace*{5mm}
{\LARGE through non-critical string\\ }
\vspace*{12mm}
{\large Kazuo Ghoroku\footnote{\tt gouroku@dontaku.fit.ac.jp}}\\
\vspace*{2mm}

\vspace*{2mm}

\vspace*{4mm}
{\large Fukuoka Institute of Technology, Wajiro, Higashi-ku}\\
{\large Fukuoka 811-0295, Japan\\}
\vspace*{10mm}
\end{center}

\begin{abstract}
We consider two-dimensional nonlinear sigma model from the viewpoint of
the holography, which has been applied to the study of the
Yang-Mills theory,
based on the non-critical string theory. We can see
the renormalization group flows 
for both the nonlinear sigma model and the Yang-Mills theory at the same time,
and the two theories
are intimately related through a kind of dual relation
of the coupling constants. We address the running behaviors of these
coupling constants and also the asymptotic freedom of the
Yang-Mills theory.

\end{abstract}
\end{titlepage}

\section{Introduction}

~~Following the conjecture of the duality \cite{M1,GKP1,W1} 
between the super Yang-Mills theory and
super-string theory, many people approached to
the Yang-Mills theory in the context of the string theory.
And an approach to the non-supersymmetric Yang-Mills theory
has been proposed in \cite{Poly1} by using
the non-critical string theory based on
the supersymetric Liouville theory. In this theory, the space-time fermions
are removed by taking the GSO projection as in the type 0
string theory which has no space-time supersymmetry \cite{DH,SW}. 
Instead, two kinds of RR-fields and
tachyon-field appear in the bulk space
other than the usual bosons of the type II string theory. While the tachyon
in the world volume of the D-brane can be removed
by imposing the GSO projection on the
open string sector. Then
the Yang-Mills field and some scalar fields are retained on the
D-brane world volume.
This formulation could be extendable to any dimension
smaller than ten, and the five-dimensional case has been
considered previously to study the pure 
4d Yang-Mills theory \cite{FM,FM2,AFS,gh1}.

On the other hand, it could be possible to identify the world-sheet action of
the non-critical string with
the fully quantized 2d sigma 
model coupled to the quantum gravity \cite{am}. 
The critical
behaviors of the sigma model are obtained from the classical
solutions of the low-energy effective action of the non-critical string
by noticing the relation of the Liouville field and the mass-scale of 
the 2d sigma model. If this approach to the 2d sigma model is 
successful, then
it would be possible to
determine the critical behaviors of two different theories, 
the Yang-Mills theory and 2d sigma model, simultaneously by one classical
solution of the non-critical string theory with D-brane background. 
The Yang-Mills fields live in the bulk space where D-branes are 
accumulated, and the sigma model is defined on the world-sheet of the
non-critical string. These
two theories, which live on different dimensional and independent
world, are connected by a higher
dimensional gravitational theory or a closed string theory.
\footnote{
An approach to this problem has recently been proposed in \cite{Schmid2} 
from the
viewpoint of the confromal field theory by exploiting linear
renormalization group equations near the conformal limit.}
The 2d sigma model has been examined by a motivation that its
properties are similar to the 4d Yang-Mills theory, i.e.
asymptotic freedom and instantons e.t.c.. From the viewpoint of the 
non-critical string theory, they are related more intimately. They 
could be studied
as a dynamically coupled system through the classical equations, which
represent the renormalization group equations of the coupled system,
of the non-critical string theory.

Previously, the
bosonic case of the non-critical string has been considered under the
linear dilaton background \cite{am}, 
but the discussions were formal since
the central charge is larger than one for the sigma model.
The problem to be resolved is the possible instability of
the tachyon in the bulk. This problem could be however
evaded by considering an appropriate potential 
\cite{CST,gho} 
or a curved space-time \cite{Bre}. When we consider the non-critical string
based on the supersymmetric world-sheet action, the RR-fields should be
added to the condensed background fields, and the possibility to
resolve the tachyon instability
is extended \cite{KT}. At least, this problem
is resolved by considering the AdS bulk space generated by
the condensation of RR-fields.
More important thing provided by the RR-fields is the existence of
the Yang-Mills theory on the boundary of this curved space.
For these two reasons, it would be meaningful to consider the non-critical
string based on the supersymmetric world-sheet action.

Our purpose is to address the critical behaviours of both theories,
the Yang-Mills theory and 2d sigma model, through
the non-critical string theory constructed by the supersymmetric
world-sheet action. The mass scale of the running coupling constants for 
both theories is given by the Liouville field.
The solutions of the low-energy
effective action are obtained as functions of Liouville field 
in order to obtain the renormalization group
equations of the two theories.

%%%%%%%%%%%%%%%%%%%%%%%%%%%%%%%%%%%%%%%%%%%%%%%%%%%%%%%%%%%%%%%%%%%%

%%%%%%%%%%%%%%%%%%%%%%%%%%%%%%%%%%%%%%%%%%%%%%%%%%%%%%%%%%%%%%%%%%%%%%
In the next section, we briefly review the relation of the 2d sigma model
coupled to quantum gravity and the non-critical string theory. And
the role of the Liouville field as a mass scale is also reviewed.
The low-energy effective action and 
the gravitational equations to be solved are given in the section
three. In section four, the critical behaviours of the sigma model
are addressed in the linear dilaton vacuum, where D-branes are
absent, according to the equations given in the previous section.
When D-branes are added, the Yang-Mills theory appears and couples to the
sigma model. In section five, we consider 
the fixed points in this case and finite coupling
constants are given. A kind of dual relation of the coupling constants
for the two theories
is shown. In obtaining these solutions, the RR-fields are
essential. Further, the running behaviours of the coupling
constants of both theories are addressed.
In section six,
we show several solutions with asymptotic freedom for either theory.
In these cases, the $d+1$ dimensional space deviates 
largely from the $AdS_{d+1}$
which appears near the horizon of the D-branes in type IIB string model.
In the final section, concluding remarks and discussions are given.

\section{2d sigma model coupled to gravity and mass scale}

Here, we briefly review a method to obtain
the renormalization group equation of the sigma model coupled to quantum
gravity in two dimension according to \cite{kpz,ddk,am}.

First, we consider a theory before coupling to quantum gravity
and a simple derivation of the
$\beta$-function in terms of conformal field theory \cite{cardy}.
Let $S_0$ denote the action of the conformal invariant theory,
and let $J$ denote a marginal operator. The total theory will be
written by:
\beq
S = S_0 + f \int d^2 z J ,\label{j1a}
\eeq
and this system will deviate away from the conformal
fixed point given by $S_0$.
Denote the short distance cut-off $\ha$, and 
assume that $J$ has the operator product expansion,
i.e.
\beq
J(r) J(0) \sim \frac{c}{r^2} J(0),~~~~r \to 0. \label{j2a}
\eeq
This implies
\beq
f^n\int d^2 z_1 d^2z_2 \la \cdots J (z_1) J (z_2) \cdots\ra_{S_0} \sim
-2\pi c\cdot \log \ha f^n \int d^2 z \la \cdots  J (z) \cdots \ra_{S_0}.
\label{ja4}
\eeq
A change in cut-off $\ha \to \ha(1+dl)$ can be compensated by
a change $f \to f -dl  \pi c f^2$ in the lower
order term of the perturbative expansion and this leads to the $\beta$-function
\beq
\beta (f) = -\frac{df}{dl} = \pi cf^2+O(f^3). \label{j5a}
\eeq

Next, let us consider the situation where the theory is coupled to
quantum gravity. We work in conformal gauge.
The metric is decomposed in a fiducial background
metric and the Liouville field $\rho$ by
\beq
g_{\a\b} = \hat{g}_{\a\b} e^{2\rho}.  \label{j6}
\eeq
At the conformal point, where $f =0$, the coupled theory
is given in Ref. \cite{ddk} as the sum of the Liouville action and
$S_0$ described by $\hat{g}_{\a\b}$.
The Liouville part of the theory can be written as \cite{ddk,ddk2}
\beq
S_L =
\frac{1}{8\pi} \int d^2z \sqrt{\hg} \left(\hg^{\a\b}\partial_\a\rho\partial_\b 
\rho -
Q\hat{R}\rho + \mu e^{\alpha \rho}\right), \label{j7}
\eeq
where
\beq
  \alpha=\left\{ \begin{array}{ll}
       - {1\over 2}(Q-\sqrt{Q^2-8}) \quad , Q=\sqrt{{25-c \over 3}}
           & \mbox{for  Bosonic} \\ 
      - {1\over 2}(Q-\sqrt{Q^2-4}) \quad , Q=\sqrt{{9-c \over 2}}
           & \mbox{for Supersymmetric} .
   \end{array}\right. 
\eeq
There are two ultra-violet cut-off: $\ha$ defined
in terms of the fiducial metric $\hg_{\a\b}$ and the physical cut-off $a$
defined by
\beq
ds^2 = e^{\a \rho} \hg_{\a\b}dz^\a dz^\b > a^2. \label{j8}
\eeq
The theory must be independent of the cut-off $\ha$ since the fiducial
metric is arbitrary: the $\b$-function must vanish.
If we consider the theory defined  by (\ref{j1a}) 
($f \neq 0$ case) which has
a non-vanishing $\b$-function before coupling to
gravity, we have a dependence of $\ha$. Then
this dependence should be eliminated by introducing new couplings
between the Liouville field and the matter fields such that
all cut-off dependence of
$\ha$ cancels order by order of $f$.
In this way, the fully quantized action can be obtained for the
theory of interacting
matter fields coupled to the gravity.
%%%%%%%%%%%%%%%%%%%%%%%%%%%%%%%%

The total action $S_L+S_{\rm matter}$ can be written by a more
general form of non-linear sigma model:
\beq 
S_{eff}=
{1 \over 4\pi}\int\,d^2z\sqrt{{\hat g}}
            \left[ {1 \over 2}G_{\mu\nu}(X){\hat g}^{\a\b}
              \partial_\a X^{\mu}\partial_\b X^{\nu}
           +{\hat R}\Phi(X)+T(X) \right], \label{5}
\eeq
where the Liouville field is represented by one of $X^{\mu}$;
$X^{\mu}=(X^0,X^i)=(\rho,X^i)$. 
In the terminology
of string theory, $G_{\mu\nu}(X)$, $\Phi(X)$ and $T(X)$ represent
the metric, dilaton and tachyon, respectively, and
they are determined such that the theory is conformal invariant.
%%%%%%%%%%%%%%%%%

Although the $\b$-function in terms of the unphysical cut-off
$\ha$ is zero by construction, we can see a 
slightly modified form of the original
$\b$-function (\ref{j5a}) from the change of the action generated
by a change of the physical cut-off $a$ defined by
(\ref{j8}). 

Consider a change $a \to a(1+dl)$ of the physical cut-off. According
to (\ref{j8}) this leads to a change $\rho \to \rho +2dl/\alpha$.
The ``running'' of the coupling
constants is seen by absorbing the shift $\rho \to  \rho+2dl/\alpha$
in a redefinition of the coupling constants $f$  \cite{schmid,5}.
This requirement determines the scale-dependence of $f$ or its
$\b$-function.
To $O(f^3)$ 
the modified $\b$-function, $\b_G(f)$, defined as $-df/dl$,  is obtained,
and it is related to the $\b$-function, $\b_(f)$, 
obtained before coupling to gravity as
\beq
\b_G (f) = \frac{2}{\alpha Q} \b (f). \label{j16}
\eeq
The modification by the pre-factor $\frac{2}{\alpha Q}$ is called
as ``gravitational dressing'' of the $\b$ function.
This is also seen in \cite{schmid,5} for $O(N)$ non-linear sigma model and
sine-Gordon model.

A more efficient way to obtain the fully quantized action 
or the renormalization group equations is to
consider in the framework of the non-critical string.
The total action $S_{eff}$ given in (\ref{5}) can be interpreted as the
world-sheet action of the non-critical string in the background
specified by $G_{\mu\nu}(X)$, $\Phi(X)$ and $T(X)$, which 
are given as a solution of classical equations of the
low-energy effective action of the string theory.
Here, the solutions should be obtained as 
\beq
  G_{ij}(X)=G_{ij}(\rho,X^i), \quad \Phi(X)=\Phi(\rho), \quad
       T(X)=T(\rho),
\eeq
and $G_{ij}(0,X^i)$ represents the original part of the sigma model.
In other words, the solutions are restricted to the form which could
represent the renormalization group equations of the coupling constants
of the original field theory coupled to the quantum gravity. 
The mass-scale of the renormalization
group is given here by the Liouville field $\rho$.

If this procedure is successful 
to see the critical behaviour of the 2d sigma model, this would be
regarded as a kind of holography. Namely,
the renormalization group flows of
2d sigma model could be found from higher dimensional gravity.
Further analysis in this context is given in section four in the linear
dilaton vacuum with non-trivial potential of the tachyon.

We consider this program in a more interesting case, where
D-branes are contained in the bulk configurations. In this case,
the gravity in the bulk is the dual of the Yang-Mills theory on the branes and
$e^{\Phi}$ represents the coupling constant of the Yang-Mills theory.
Then we can see the running of both coupling constants simultaneously
by solving the same gravitational equations under the assumption that
$\rho$ represents the mass scale for both field theories.
One more merit to consider D-branes is that
it would be possible to extend the sigma model to
the case of $c>1$ since the background has a curvature to stabilize
the tachyon.

%s1s%%%%%%%%%%%%%%%%%

\section{Effective action and equations to be solved}

Let us consider here the super-symmetric case of the non-linear sigma model
which has $d+n=D-1$ boson fields; The boson part of the action is written
as 
\beq
S = \frac{1}{8\pi} \int d^2 z\sqrt{g}
\left\{ \sum_{\mu=1}^d \partial_\b \phi^\mu \partial^\b \phi^\mu +
\sum_{i,j=1}^n \hat{G}_{ij}(\phi) \partial_\b \phi^i \partial^\b \phi^j
\right\} , \label{31}
\eeq
where ${\hat G}_{ij}(\phi)$ denotes the $S^n$ metric, and the fermionic parts
are neglected since they are not used here.
The full super-symmetric
and general form of
action would be found in \cite{FT}. This model consists of $d$-free fields
and $n$ interacting fields, which form the $O(n+1)$ non-linear sigma model 
here. The sector of the free fields are prepared for the space-time of the
D-branes where Yang-Mills fields are living. The interesting case is $d=4$,
but we consider in general $d$ here.
In the 2d world-sheet action the parameters of the
Yang-Mills theory can not be seen until the theory is coupled
to the quantum gravity. 
When the model is coupled to 2d quantum gravity and quantized
in the conformal gauge,
we get (ignoring as usual the ghost terms) the 
following boson part of the quantized
action as
\bea
S(\hg) &=& \frac{1}{8\pi} \int d^2 z\sqrt{\hg}
\Bigg\{ G_{00}(\rho)\partial_\b \rho \partial^\b \rho + \hat{R}\Phi(\rho)
 +G_{MN}(\rho,\phi)\partial_\b \phi^M \partial^\b \phi^N  \nonumber \\ 
    &{}&  \qquad\qquad~~~~~~ + (RR-\rm{bg}) \Bigg\}, \label{smg1}
\eea
where $M, N = 1 \sim D-1$, 
$G_{MN}(\rho,\phi)=\{ G_{\mu\nu}(\rho), G_{ij}(\rho,\phi)\}$  and 
$\rho$ denotes the Liouville field. 
The last term $(RR-\rm{bg})$ represents the part depending of the 
$RR$-background fields which appear since we are considering super-symmetric
world sheet action of the corresponding string model.
$\Phi(\rho)$, $G_{00}(\rho)$, $G_{MN}(\rho,\phi)$ and $RR- \rm{bg}$
are determined such that the above action is
conformal invariant as mentioned in the previous section.
From the viewpoint of the holography, we comment
on these functions: (i) When the D-branes are present, 
$e^{\Phi(\rho)}$ represents the Yang-Mills coupling constant.
(ii) As for $G_{MN}(\rho,\phi)$, its sigma model part can be denoted as 
$G_{ij}(\rho,\phi)=e^{2C(\rho)}\hat{G}_{ij}(\phi)$
according to the principle mentioned in the previous section, and 
$e^{2C(\rho)}$ represents the running coupling constant of the sigma model.
We must notice here that it is controlled by the Yang-Mills theory or D-branes.
The free
field part is also non-trivial in the presence of D-branes, and it
is denoted here as $G_{\mu\nu}(\rho)=e^{2A(\rho)}\eta_{\mu\nu}$.
This is the reflection of the fact that the sector of the Yang-Mills theory
interacts with the sigma model sector via Liouville fields $\rho$.

This action represents the boson part of the super-symmetric 
world-sheet action of the non-critical string which couples to the background
defined by $G_{MN}(\rho,\phi)$, $\Phi(\rho)$ and $RR-bg$. Here we suppose 
a string theory of type 0 in which there is no space-time super-symmetry
but modular invariance of the loop amplitudes is satisfied.
In the target-space, only one R-R field
$A_{p+1}$ is considered other than the usual NS-NS fields.
\footnote{Although two kinds of the fundamental D-brane can be considered
\cite{KT}, we here consider only the electric type of R-R charge.
} 
And we expect
that N $D_{p+1}$-branes are stacked on the boundary to make
the U(N) gauge theory there. 

Then we start from the following target-space action,
\bea
    S_D &=& {1\over 2\kappa^2}
      \int dx^D\sqrt{|g|}\Bigg\{e^{-2\Phi}\left(R-
      4(\nabla\Phi)^2+(\nabla T)^2+V(T)+c \right) \nonumber \\ 
    &{}&  \qquad\qquad~~~~~~ + \frac{1}{2(p+2)!}f(T)F_{p+2}^2\Bigg\},
    \label{action} 
\eea
where $c=-(10-D)/2\alpha'$, and $F_{p+2}=dA_{p+1}$ is the 
field strength of $A_{p+1}$. Hereafter we take $\alpha'=1$.
The total dimension
$D$ includes the Liouville direction, which was denoted by $\rho$.
The tachyon potential is represented by
$V(T)$, and $f(T)$ denotes the couplings between the tachyon and the R-R
field investigated in \cite{KT}. 
When we solve the equations near $T=0$, we use for 
$V_c(T)=c+V(T)$ its well-known form,
\beq
   V_c= {D-10\over 2}- T^2+ O(T^4) \,  . \label{pot}
\eeq
In the case of super-symmetric world-sheet action, the potential would be the
even function of $T$ \cite{KT}.

Although $f(T)$ is expected to play an important role in the type 
0 model \cite{KT},
we neglect here its $T$-dependence and set $f(T)=1$ for the simplicity.
As for the stability of the tachyon in the dimension $D>2$,
it is recovered here by $V(T)$ and D-branes as seen below. Then
the equations of motion are written as
\bea
     R_{\mu\nu} - 2 \nabla_\mu \nabla_\nu \Phi &=& -\nabla_\mu T\nabla_\nu T + 
             e^{2\Phi} T_{\mu\nu}^A \label{metric}\\
    -4 \nabla_\mu \Phi \nabla^\mu \Phi 
       +2 \nabla^2 \Phi &=& \frac{D-2d-2}{4(p+2)!}e^{2\Phi} F_{p+2}^2
            + V_c(T) \label{dilaton}\\
    \nabla^2 T -2 \nabla_\mu \Phi \nabla^\mu T 
         &=& \frac{1}{2}V'_c(T)  \,
            \label{tachyon}
\eea
\beq
     \partial_\mu (\sqrt{|g|} F^{\mu\indi}) = 0 \, , \label{form}
\eeq
where 
\beq  
 T_{\mu\nu}^A = -\frac{1}{2(p+1)!}
                \left(F_{\mu\indi}F_\nu^{\;\;\;\;\indi}
       - \frac{g_{\mu\nu}}{2(p+2)}  F_{\indii}F^{\indii}\right).
     \label{tensor3} 
\eeq
We solve the above equations according to the following ansatz;
\beq
 ds^2= e^{2A(r)}\eta_{\mu\nu}dx^{\mu}dx^{\nu}
           +e^{2B(r)}dr^2 +e^{2C(r)}\hat{G}_{ab}dx^adx^b
         \, \label{metrica}
\eeq
\beq
    \Phi\equiv\Phi(r), \quad \quad T\equiv T(r) \quad \hbox{and} \quad
     A_{01\cdots p} = -e^{q(r)}\ ,
\eeq
where $(x^{\mu},x^a)$, $\mu=0\sim p(=d-1)$ and $a=1\sim n$, 
denote the space-time coordinates.
And $r$ is related to the Liouville direction as given below.
The equation (\ref{form}) is solved as
\beq
 \partial_{r}e^{c(r)}=N e^{dA+B+\bar{d}C} \, \label{RRfi}
\eeq
where $\bar{d}=D-d-1=D-p-2$ and N denotes the number of the p-brane.
The remaining equations (\ref{metric}) and (\ref{dilaton}) are 
solved in the form
\beq
  A(\rho)=\gamma\rho+a(\rho), \qquad B(\rho)=-\rho+b(\rho), \label{an1}
\eeq
\beq
    C(\rho)=C_0+c(\rho), \quad
    \Phi(\rho)=\Phi_0+\phi(\rho), \quad T(\rho)=T_0+t(\rho), \label{an2}
\eeq
where we used the following notation for the Liouville mode,
          $$\rho=\ln r .$$ 
This relation is given such that the flat metric of $d+1$ sector
could be found for $\gamma=0$, $a=0$ and $b=0$ when their coordinates
are taken by $(\rho, x^{\mu})$ as in considering the sigma model.

Then the equations to be solved are given as
\beq
      \ddot{a}+\dot{a}(d\dot{a}-\dot{b}+\bar{d}\dot{c}-2\dot{\phi})+
       \gamma(2d\dot{a}-\dot{b}+\bar{d}\dot{c}-2\dot{\phi}) 
         =-d\gamma^2+{\lambda_0^2\over 4}
       e^{2l}, \label{metric11}
\eeq
\beq
       d[\ddot{a}+\gamma(2\dot{a}-\dot{b})
                +\dot{a}(\dot{a}-\dot{b})]
           -2(\ddot{\phi}-\dot{b}\dot{\phi})+\dot{t}^2+
         \bar{d}[\ddot{c}+\dot{c}(\dot{c}-\dot{b})] 
          =-d\gamma^2+{\lambda_0^2\over 4}
       e^{2l}, 
                \label{metric21}
\eeq    
\beq
      \ddot{c}+\dot{c}(d\dot{a}-\dot{b}+\bar{d}\dot{c}-2\dot{\phi})+
       d \gamma \dot{c}
         =(\bar{d}-1) e^{-2C_0+2(b-c)}-{\lambda_0^2\over 4}
       e^{2l}, \label{metc}
\eeq
\beq
    \ddot{\phi}+\dot{\phi}(d\dot{a}-\dot{b}+\bar{d}\dot{c}-2\dot{\phi})
       +d\gamma\dot{\phi}
     = -{D-2d-2 \over 8}\lambda_0^2 e^{2l}
          +{1\over 2}V_c(T) e^{2b} \, ,
                 \label{dila21}
\eeq
\beq
    \ddot{t}+\dot{t}(d\dot{a}-\dot{b}+\bar{d}\dot{c}-2\dot{\phi})
      +d\gamma\dot{t} = {1\over 2} V_c'(T) e^{2b} \, ,
         \label{tac21}
\eeq
where 
   $$l=b+\phi-\bar{d}(C_0+c), \quad \lambda_0=Ne^{\Phi_0}.$$
The dot denotes the derivative with respect to $\rho$, and
$\lambda_0$ represents the t'Hooft coupling constant.
%%%%%%%%%%%%%%%%%%%%%%%%%%%%%%%%%

\section{Solution in linear dilaton vacuum }

Here we firstly examine
the renormalization group equations of the sigma model only
before studying the coupled system of the Yang-Mills theory
and the sigma model. The equations are obtained 
from the above equations by setting as $N=0$, which means no D-branes
in the background. In this case, the simplest configuration is the
linear dilaton vacuum, which is set as
\beq
   \gamma=0 \, , \quad 
     \phi(\rho)=-{1\over 2}Q\rho+\varphi(\rho). \label{ld1}
\eeq
Now the d-dimensional part is not necessary since the D-branes are
absent, then we consider the case $d=0$ for the simplicity.
Further, we take $t(\rho)=0$, which represents the fluctuation of $T$
around $T=T_0$, and $T_0$ is specified by
\beq
  V'(T_0)=0 .
\eeq
Then the functions to be solved are $b, c$
and $\varphi$. Their equations are obtained from
(\ref{metric21}) $\sim$ (\ref{dila21}) as follows:
\beq
     \bar{d}[\ddot{c}+\dot{c}(\dot{c}-\dot{b})] -Q\dot{b}
          = 2(\ddot{\varphi}-\dot{b}\dot{\varphi}) , 
                \label{metric22}
\eeq    
\beq
      \ddot{c}+\dot{c}(-\dot{b}+\bar{d}\dot{c}-2\dot{\varphi})+
       Q \dot{c}
         =(\bar{d}-1) e^{-2C_0+2(b-c)}, \label{metc2}
\eeq
\beq
    \ddot{\varphi}+\dot{\varphi}(-\dot{b}+\bar{d}\dot{c}-2\dot{\varphi})
       +Q(2\dot{\varphi}-{1\over 2}\bar{d}\dot{c})
     = {1\over 2}[Q^2+V_c(T_0)e^{2b}]  \, .
                 \label{dila22}
\eeq

The fixed point of the sigma model is easily seen by setting $\dot{c}=0$
in Eq.(\ref{metc2}). Then we find the fixed point at,
\beq
  g_N=e^{-2C}=0. \quad
\eeq
Further, we should take $\dot{\phi}=0$ since the linear dilaton vacuum
is considered as the fixed point. Then 
the background charge is obtained from Eq.(\ref{dila22})
as $Q=e^{b^*}\sqrt{-V_c(T_0)}$, $V_c(T_0)<0$ is required. And
$b^*$ is a constant as seen from Eq.(\ref{metric22}).

We can check the stability for the tachyon fluctuation at this fixed
point, where the linearized 
equation of (\ref{tac21}) is written as,
\beq
    \ddot{t}+Q\dot{t} = {1\over 2}V_c''(T_0)t  \, ,
         \label{tacd}
\eeq
where $V_c'(T_0)=0$ is used. Then we obtain the following condition
of stability,
\beq
     V_c''(T_0)\geq {D-10 \over 4} \, .
         \label{taco}
\eeq
When we consider the case $T_0=0$, which leads to $V_c''(T_0=0)=-2$, 
the above condition is reduced
to $D\leq 2$, which is the well-known result. 
If we consider to extend this fixed point up to
$D=10$, then we must require $T_0\neq 0$ and $V_c''(T_0)\geq 0$. 
%%%%%%%%%%%%%%%%%%%%%%%%%
%===  Fig.  ========================================================
\begin{figure}[htbp]
\begin{center}
   \includegraphics[width=7cm,height=7cm]{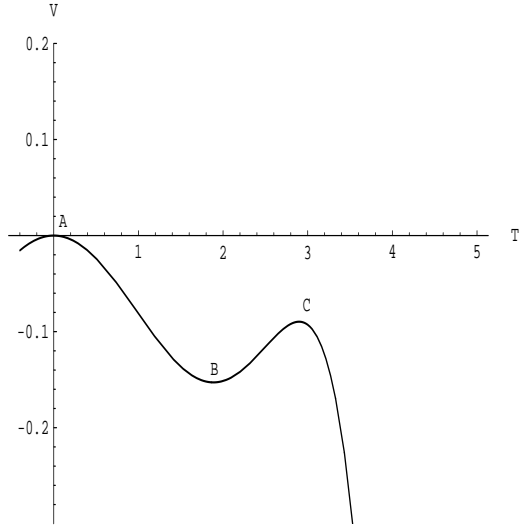}
 \caption{A schematic graph of the tachyon potential $V(T)$, where
the scale is arbitrary. The origin {\bf A} represents the perturbative
vacuum. The minimum point {\bf B} corresponds to the linear dilaton
vacuum, and 
{\bf C} is the point where the asymptotic free Yang-Mills theory with
$\gamma\neq 0$ is realized.}
\end{center}
\end{figure}
%====================================================================

The possible position
of this vacuum is shown in Fig.1, which shows the schematic
tachyon-potential since the correct form of $V(T)$ is not known. 
The point $B$ represents
the minimum of $V(T)$, and such a point will be found if the
second order term of $T^4$ is positive. But this is open here,
and we assume that there is such a point.

Now, we study the running behaviour of $g_N$ near this fixed
point. The running solutions are obtained by expanding 
$\varphi, b$ and $c$ around this fixed point by an appropriate small
parameter. Here two expansion-forms are considered.
First, we expand the running parameters by $g_N$
as in the usual perturbation since we are considering near $g_N=0$.
Then $\varphi$ and $b$ are expanded as
\beq
 \varphi=\varphi_1g_N+\varphi_2g_N^2+\cdots, \quad
 b=b_1g_N+b_2g_N^2+\cdots.   
\eeq
And the $\beta$-function of $g_N$ is defined as 
\beq
    \beta_g=\dot{g_N}=\beta_0g_N^2+\beta_1g_N^3+\cdots ,
\eeq
since the loop-corrections for the coupling constant
begins from $O(g_N^2)$. Then the equations
(\ref{metric22}) $\sim$ (\ref{dila22}) are rewritten in terms of $\beta_g$,
for example (\ref{metc2}) can be written as
\beq
 \beta[g_N\beta'+Qg_N-\beta({\bar{d}\over 2}+1+2g_N\varphi'+g_Nb')]
         =-2g_N^3(\bar{d}-1)e^{2b} , \label{metc22}
\eeq
where prime denotes the derivative with respect to $g_N$.
From the lower order equations, we obtain
\beq
 \beta_0=-2{\bar{d}-1 \over Q}, \quad
  \beta_1=-4{(\bar{d}-1)^2\over Q^3}, \label{beta1}
\eeq
\beq
 b_1={\bar{d}(\bar{d}-1) \over 2Q}, \quad
  \varphi_1=-{5\bar{d}(\bar{d}-1)\over 16Q^2}+{\bar{d}\over 8}.
\eeq
The "dressed" $\beta$-function should be defined as 
$\beta_g=dg_N/\alpha d\rho$, here $\alpha=-1/2(Q-\sqrt{Q^2+2V''(T_0)})$,
according to the section two. Then we obtain
\beq
 \beta_g={2\over \alpha Q}(-(\bar{d}-1)g_N^2-
              2{(\bar{d}-1)^2\over Q^2}g_N^3+\cdots). \label{betag}
\eeq
This is the expected $\beta$-function of $g_N$ \cite{am} since
it shows the asymptotic freedom and the dressed factor given in (\ref{j16}).
The coefficient is exact at least for the one loop result. But the 
coefficient of the two-loop correction depends of $Q$ even if the
dressed factor was moved out. 
This implies that the relation given in (\ref{j16})
is not correct at two-loop order and the dressed factors are different
order by order. The second remark is that the $\beta$-function given above
is considered as
the one of super-symmetric O$(\bar{d})$-nonlinear sigma model. And
it is known that the $\beta$-function is one-loop finite for O$(3)$
super-symmetric sigma model \cite{NSVZ}. But this can not be seen from
(\ref{betag}) when we respect to the relation (\ref{j16}), so we should 
consider that the second order term would be 
largely modified by the quantum gravity.
However the one-loop part is unchanged and Eq.(\ref{j16}) is satisfied.
Noticing this point, we continue the analysis further.

Next, we solve the same equations, (\ref{metric22}) $\sim$ (\ref{dila22}),
in terms of another expansions. The asymptotic free coupling constant
generally has the behaviour like $1/\rho^a$ with positive constant $a$
near the fixed point, where $\rho\to 0$. 
Then we expect $c=c_0\ln \rho+\cdots$, where $\cdots$
tends to zero for $\rho\to \infty$ and $c_0>0$. Then by assuming the form,
\beq
 C=C_0+c_0\ln\rho+\cdots, \quad b=b_0\ln\rho+\cdots, \quad
  \varphi=\varphi_0\ln\rho+\cdots ,  \label{sing}
\eeq
we obtain the following solution,
\beq
 c_0={1\over 2}, \quad b_0=0, \quad \varphi_0={\bar{d}\over 8}, 
        \quad \rm{and} \quad
   e^{2C_0}={2(\bar{d}-1)\over Q} .
\eeq
This result gives the same $\beta_g$ up to the one loop order, and we
can see $g_N\sim 1/\rho$ at large $\rho$. But the $\varphi(\rho)$, which 
begins from $\ln\rho$ here, has
different behaviour from the one obtained by the $g_N$-expansions,
where $\varphi(\rho)$ begins from $1/\rho$. 
Both expansions are the
perturbative expansions around the same fixed point, so the differences
in the $\rho$ dependence of the parameters of the theory could be considered
as the difference of the regularization scheme, but the one-loop
coefficient of the $\beta$-function is scheme independent. 
We notice here that $\varphi(\rho)$ does not now represent the coupling 
constant of the gauge theory, which is absent in this case.

%%%%%%%%%%%%

The above equations, (\ref{metric22}) $\sim$ (\ref{dila22}),
might be used even at large coupling region. And
we would expect an infrared fixed point at some
point $g_N=g_N^*\neq 0$, which smoothly continued to the ultra-violet
fixed point at $g_N=0$ given above. If such a fixed point existed, it
might be found by using the second type expansions,
\beq
 \beta=\beta_0(g_N-g_N^*)^{\beta_1}+\cdots , \quad \beta_1>0
\eeq
\beq
 b=b_0\ln (g_N-g_N^*)+\cdots , \quad 
     \varphi=\varphi_0\ln (g_N-g_N^*)+\cdots .
\eeq
But we could not find such a solution. Then we could say
that the sigma model has only the ultra-violet fixed point at $g_N=0$.

In the following sections, we consider the case of $\lambda_0\neq 0$,
where the D-branes are in the
background and $\phi(\rho)$ represents the gauge coupling constant. And
we see how the above critical behaviors of the sigma model will
be changed.

%%%%%%%%%%%%%%%%%%%%%%%%%%% Linear Dilaton %%%%%%%%%%%%%%%%%%%5

\section{Solution in background with D-branes}

Here we consider the equations, (\ref{metric11}) $\sim$ (\ref{tac21}),
for the case of $N\neq 0$ and $\gamma\neq 0$. Then the terms of the D-branes
are included in the equations. The fixed point of $g_N$ and 
$\lambda=Ne^{\Phi}$ is simply seen by taking them as constants, $g_N^*$
and $\lambda^*$. Then we obtain the following relations from Eqs.(\ref{metc})
and (\ref{dila21}) by neglecting their left hand sides,
\beq
       (\bar{d}-1) g_N^*  = {\lambda^2 \over 4}g_N^{*\bar{d}}
       , \label{metc0}
\eeq
\beq
     (D-2d-2){\lambda^2 \over 4}g_N^{*\bar{d}}
               = V_c(T^*)  \, . \label{dila0}
\eeq
From these, we obtain
\beq
   \lambda^{*2} g_N^{*\tilde{d}} = 4\tilde{d}, \label{coupl}
\eeq
where $\tilde{d}=\bar{d}-1=D-d-2$. We consider the case of $\tilde{d}>0$.
The relation (\ref{coupl}) implies a kind of duality of
$\lambda^*$ and $g_N^*$. The asymptotic freedom of gauge coupling constant
could be seen in the large coupling region of $g_N$ and vice versa. 
This relation would be satisfied in quite general $d+1$-dimensional manifold
since Eq.(\ref{coupl}) is independent on $a(\rho), b(\rho)$ and $\gamma$ which
are responsible for the geometry of the $d+1$-dimensional manifold.
It is possible to determine the geometry of the background manifold for
the fixed point satisfied (\ref{coupl}) by solving Eqs.(\ref{metric11}) and
(\ref{metric21}) with respect to $a(\rho)$ and $b(\rho)$. By assuming
$t=0$, we can solve them as
\beq
 \dot{a}={1\over 4}\lambda^{*2}g_N^{*\bar{d}}e^b-\gamma ,
\eeq
where $b(\rho)$ is arbitrary. When $b$ is remained as unknown function of 
$\rho$, the analysis below becomes very obscure. Then we make an anzatz that
$a(=a^*)$ and $b(=b^*)$ are constant at the fixed point also.
By taking further as $\dot{t}=0$, we obtain from the remaining equations
\beq
         d\gamma^2 = {\lambda^2 \over 4}g_N^{*\bar{d}}e^{2b^*}
       , \label{met0}
\eeq
\beq
      V_c'(T^*) = 0 \, .
         \label{tac0}
\eeq
These results are obtained by neglecting the $\rho$-dependence of all
fluctuations,
$\chi_i=$ $\{$ $a(\rho), b(\rho), \phi(\rho), t(\rho)$ $\}$,
from the biggining. And we arrive at the fixed point configuration,
$AdS_{d+1}\otimes S^n$.

We firstly
solve these near $T=0$, then $T^*=T_0=0$ is obtained from Eqs.(\ref{tac0}) 
and (\ref{pot}).
This corresponds to the point $A$ in Fig.1 of the tachyon potential.
Other parameters are determined by solving the remaining equations. Possible
points examined here are shown in $D-d$ plane of Fig.2,
but all these points 
are not stable. The stability condition is obtained by solving the linearized 
equation of Eq.(\ref{tac21}) with respect to $t$ such that it has
a real solution in the form of $t=e^{\alpha\rho}$, i.e.
$\alpha$ being real. The equation is given as
\beq
  \ddot{t}+d\gamma\dot{t} = -te^{2b^*} , \label{tac2}
\eeq
then the requirement is expressed by,
\beq
        (d\gamma/e^{b^*})^2\geq 4 , \label{rel1}
\eeq 
since $\alpha$ is obtained as 
\beq
  \alpha={1\over 2}(-d\gamma\pm\sqrt{(d\gamma)^2-4e^{2b^*}})  . \label{al}
\eeq
After all, we find three stable solutions which satisfy the above
condition (\ref{rel1}), and they are shown by disks in the Fig.2. 
The parameters for these solutions are summarized in the Table 1.

%===  Fig.  ========================================================
\begin{figure}[htbp]
\begin{center}
   \includegraphics[width=7cm,height=7cm]{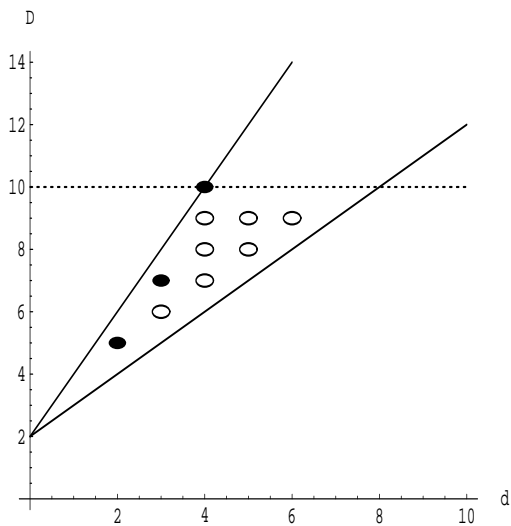}
 \caption{The circles are unstable against tachyon perturbation
and the disks are the stable fixed points. The lines represent
(a) $D=2(d+1)$ and (b) $D=d+2$.}
\end{center}
\end{figure}
%====================================================================

%%%%%%%%%%%%%%%%%%%%%%%%% Table %%%%%%%%%%%%%%%%%%%%%%%%%%%%%%%%%

\begin{center}
$$\begin{array}{|c|c|c|c|c|c|}\hline
 \mbox{} & \mbox{(D,d)}  & \mbox{$\lambda^*$} & \mbox{$g_N^*$} 
                  & \mbox{$\gamma/e^{b^*}$}  & \mbox{n}   \\ \hline
 \mbox{sol.1} & \mbox{(10,4)}  & \mbox{$4(e^{b^*}/\gamma)^4$} 
            & \mbox{$(\gamma/e^{b^*})^2$} 
                  & \mbox{any constant $\geq 1/2$}  & \mbox{5}   \\ \hline
 \mbox{sol.2} & \mbox{(7,3)}  & \mbox{$8\sqrt{2}/3$} & \mbox{${3\over 4}$} 
                  & \mbox{${1\over\sqrt{2}}$} & \mbox{3} \\ \hline
 \mbox{sol.3} & \mbox{(5,2)}  & \mbox{$2\sqrt{2/5}$} & \mbox{${5\over 2}$} 
                  & \mbox{${\sqrt{5}\over 2}$} & \mbox{2}  \\ \hline
\end{array}$$
Table 1. The stable fixed points.
\end{center}
%%%%%%%%%%%%%%%%%%%%%%%%%  Table %%%%%%%%%%%%%%%%%%%%%%%%%%%%%%%%%%%%%%%%%%%%

We give several comments on these solutions:
The (sol.1) is special since 
Eq.(\ref{dila0}) does not give any constraint in this case, and all the values
of parameters are not determined. But they are bounded from the stability
as, $g_N^*\geq 1/4$ and $\lambda^*\leq 64$. And we can see the relation
of these two critical couplings,
\beq
  g_N^{*2} \lambda^* = 4 , \label{coupl2}
\eeq
which is independent of $\gamma/b^*$. Although $g_N^*$ approaches to zero with
decreasing $\gamma$, it is however bounded from below due to the stability and
it can not arrive at zero. Then it would be impossible to
see the asymptotic free region of the sigma model in the background
with D-branes and $T_0=0$. This can be interpreted as
that the fixed point observed at $g_N=0$ 
in the linear dilaton vacuum was pushed to the finite value 
by D-branes or Yang-Mills theory.  If $\gamma$ decreased across the lower
bound, the the tachyon would condense to
stabilize the system and a new fixed point will be found in the different
background where $T_0\neq 0$.

On the other hand, it will be possible to consider the limit of $\lambda^*=0$
in this vacuum, and this is
the asymptotic free limit of the gauge theory.
In this case, the equations for 
the running couplings could be solved by expanding them in the series
of $\lambda$ as done in the previous section for $g_N$. 
In performing this expansion,
we should notice $g_N^*\to \infty$ and $e^{b^*}\to 0$ for $\lambda^*\to 0$.
Then we parametrize them as
\beq
 b(\rho)=b_0\ln \rho +b_1/\rho+\cdots, \quad
  c(\rho)=c_0\ln \rho +c_1/\rho+\cdots,
\eeq
and introduce $\beta_{\lambda}=d\lambda/d\rho$ which is expected to
be expanded as $\beta_{\lambda}=\beta_0\lambda^3+\cdots$. However we
can not find any solution in this form. The reason why this expansion
has failed would be reduced to the fact that the right hand side of 
Eq.(\ref{dila21}) vanishes. But the terms of the right hand side
are necessary to obtain a desired
renormalization group flow of the Yang-Mills theory.
%%%%%%%%%%%%%%
This point can be understood from the solution obtained in the previous 
section, where the first term of the right hand side
of Eq.(\ref{metc}) was necessary to obtain the expected asymptotic free
$\beta$-function of the sigma model. In fact also for $\lambda$,
the asymptotic free solutions 
are seen by using the above expansion in the next section where the 
second term of the right hand side of Eq.(\ref{dila21}) is retained by
considering the vacuum with $T_0\neq 0$.
%%%%%%%%%%%%
Then the running behaviours of the solution (sol.1) can be seen 
in a different expansion-form which is suitable
to the case of the finite coupling constant.

For solutions (sol.2,3), the parameters are all
determined definitely. The value of $b^*$ depends on $\gamma$, but
$\lambda^*$ and $g_N^*$ are fixed independently on $\gamma$. Then
the coupling constants at the fixed points are fixed definitely, this
is due to availability of the Eq.(\ref{dila0}). Although
the field theory on the branes are different from the 4d Yang-Mills theory, 
we can expect the similar features of the flow equations to the case
of (sol.1). As shown below, these are the ultra-violet fixed
points for both sigma model and the field theory on the branes. 

\vspace{.5cm}
\noindent{\bf Running behaviours :}
First we consider the (sol.1), where we can see
the relation of the sigma model and the 4d Yang-Mills theory.
Their coupling constants at fixed points are related by Eq.(\ref{coupl2}).
Near the fixed point,
the equations can be approximated
by the linearized one;
\beq
      \ddot{a}+\gamma(2d\dot{a}-\dot{b}+\bar{d}\dot{c}-2\dot{\phi}) 
         = {\lambda_0^2\over 2}(\phi-\bar{d}c+b), \label{metl1}
\eeq
\beq
       d\ddot{a}+d\gamma(2\dot{a}-\dot{b})+\bar{d}\ddot{c}
           -2\ddot{\phi} = {\lambda_0^2\over 2}(\phi-\bar{d}c+b), 
                \label{metl2}
\eeq    
\beq
      \ddot{c}+ d \gamma \dot{c}
         =-{\lambda_0^2\over 2}[\phi-(\bar{d}-1)c], \label{metcl}
\eeq
\beq
    \ddot{\phi} +d\gamma\dot{\phi} = 0 \, ,
                 \label{dil2}
\eeq
\beq
    \ddot{t}+d\gamma\dot{t} = {1\over 2} V_c''(T_0)t \, .
         \label{tacl}
\eeq
First, we can solve Eqs.(\ref{metcl}) and (\ref{dil2}) with respect to 
$c$ and $\phi$. They represent how two dynamical system, 2d sigma model
and 4d Yang-Mills theory, are interacting. Once some non-trivial $c$ is
given, the running behaviour of $\phi$ is given from Eq.(\ref{metcl}).
In other words, the $\beta$-function of the Yang-Mills theory is induced by 
the 2d sigma model and vice versa. 
By using $\phi$ and $c$ determined by Eqs.(\ref{metcl}) and (\ref{dil2}), 
$a$ and $b$ are obtained from Eqs.(\ref{metl1}) and (\ref{metl2}).
The equation of $t$, (\ref{tacl}), is separated alone, then
it does not give any essential effect to the running couplings of two
theories.

Eq.(\ref{dil2}) is solved as $\phi=\phi_1 e^{-d_0\rho}$, where $d_0=d\gamma$
and $\phi_1$ is some constant.
Then solutions for $\chi_i$ can be solved
systematically order by order in terms of the 
following expansions,
\beq
  \chi^i(r)=\sum_{n=1}^{\infty} \chi^i_n(e^{-d_0\rho})^n \, . \label{expa}
\eeq
For example, $a(r)=\sum_n$ $a_n(e^{-d_0\rho})^n$.
Up to the second order, the coefficients $\{$ $c_1,
c_2$ $\}$ and $\{$ $a_1,a_2$ $\}$ are taken arbitrarily, and the one of
$\phi, b$ and $t$ are represented by them:
\beq
  \phi_1= 4c_1 ,\quad \phi_2= 16c_1(c_1-a_1) ; 
  \quad t(\rho)= O(e^{-3\lambda_0\rho}) , \label{coe1}
\eeq
\beq
  b_1=5c_1-4a_1 , \quad
  b_2=-8a_2-{41\over 3}c_2+{8\over 3}(29c_1^2-28c_1a_1-3a_1^2) .
 \label{coe2}
\eeq
From these perturbative results, the $\beta$-function, which is defined as
$\beta(\lambda)\equiv d\lambda/d\rho$ for $\lambda(=N e^{\Phi}=Ng_{YM}^2)$,
satisfies the following relation at the critical point:
\beq
 \beta'(\lambda^*)= -d_0=d\gamma  . \label{beta}
\eeq
For $g_N=e^{-2C}$, we obtain
$\beta'(g_N^*)= -2d_0/|\alpha|$ by re-scaling the scale parameter,
where $\alpha$ is given in (\ref{al}). 
This means that the fixed point given above is the ultra-violet one
for both 2d sigma model and the 4d Yang-Mills theory. 

\vspace{.5cm}

In the case of $D < 10$,
there are two stable solutions at $(D,d)=(7,3)$ and $(5,2)$ as shown in the
Table 1. In both cases, the d-dimensional field theories are
not the 4d Yang-Mills theory and $\Phi$ would not correspond to
the running coupling-constant of the Yang-Mills theory. But we can
see the similar relations between the sigma model 
and the theory in the branes as in the case of $D=10$.
They affect each other and the fixed points of both theories
are at finite values as given in the Table 1. As for the 
linearized equations of $\chi^i$ are the same with the 
one of $D=10$ except for Eq.(\ref{dil2}), which is written here by
\beq
    \ddot{\phi} +d\gamma\dot{\phi} = -{D-10\over 2}(\phi-\bar{d}c) \, .
                 \label{dil3}
\eeq
The equations are solved as in the case of $D=10$. 
First, this equation and Eq.(\ref{metcl}) are solved in the form of
$\phi=\phi_1e^{\delta\rho}$ and $c=c_1e^{\delta\rho}$, then we obtain
\beq
 \delta=-1\pm\sqrt{E_{\pm}}, \qquad 
   E_{\pm}={1\over 2}(D-2\pm\sqrt{D^2-12D+84}) .
\eeq
The value of $\delta$ are evaluated for the two cases as
\beq
  \delta=\left\{ \begin{array}{ll}
        -4.30,\, 2.07,\, -1.12\pm 0.39i
           & \mbox{for  (D,d)=(5,2)} \\ 
        -4.09,\, -1.81,\, -0.308, \, 1.97
                  & \mbox{for (D,d)=(7,3)} .
   \end{array}\right. 
\eeq
It should be real and negative since
we are considering at the ultra-violet fixed point. In this sense,
we can choose
$\delta=-4.30$ for $(D,d)=(5,2)$, but there are three candidates
in the case of $(D,d)=(7,3)$ and there is no principle to choose
one of them. Including this problem, these two solutions are remained
to be explained in more detail based on some meaningful dynamical system.

\section{Asymptotic free solution}

The fixed points considered above correspond to a finite value
of the coupling constants for both gauge theory and the sigma model.
However we can expect a fixed point where asymptotic freedom is seen for
one of them due to the relation (\ref{coupl}).
Such a solution would be found by solving the equations
in terms of the following expansion for $\chi^i$ 
\cite{Min2,KT3}. For $a$, we take the form,
\beq
  a(\rho)=\bar{a}_0\ln \rho+\bar{a}_1{1\over \rho}(\ln \rho+\bar{a}_{10})
         +\cdots  , \label{asyma}
\eeq
For $b$ and $c$, they are written by replacing ($a$, $\bar{a}_0$
$\bar{a}_{10}$) by ($b$, $\bar{b}_0$, $\bar{b}_{10}$) and
($c$, $\bar{c}_0$, $\bar{c}_{10}$) respectively in the above
equation (\ref{asyma}). And we set
\beq
  \phi(\rho)=\bar{\phi}_0\ln \rho+\bar{\phi}_1{1\over \rho}\ln \rho
       +\bar{\phi}_{2}{1\over \rho^2}\ln^2\rho+\cdots , \label{asymp}
\eeq
\beq
  t(\rho)= \bar{t}_1{1\over \rho}+
         \bar{t}_2{1\over \rho^2}(\ln \rho
       +\bar{t}_{20})+\cdots  . \label{asymd}
\eeq
In these functional form, the asymptotic free gauge theory is obtained 
for $\bar{\phi}_0<0$. While the asymptotic free behaviour
of the 2d sigma model will be found for $\bar{c}_0>0$.

\vspace{.5cm}
\subsection{Solutions for $\gamma\neq 0$}

First, we consider the case of $\gamma\neq 0$.
The linearized equations with respect to the fluctuations 
$\chi^i$ are not useful when we use the above expansions, and
full form of equations are now used. And they are solved by expanding 
equations in the power series of $1/\rho$, which is assumed being very 
small. Firstly the lowest order terms are examined. From Eqs.
(\ref{metric11}), (\ref{metric21}) and (\ref{metc}), 
we obtain 
\beq
  {\lambda_0^2\over 4}e^{-2\bar{d}C_0}
        =d\gamma^2=(\bar{d}-1)e^{-2C_0} , \label{AF01}
\eeq
\beq
  \bar{\phi}_0=(D-d-2)\bar{c}_0 , \quad 
     \bar{c}_0 = \bar{b}_0 . \label{AF1}
\eeq
The above solution is possible only for $D-d-2 > 0$ since we are considering 
the case $\gamma\neq 0$.
From (\ref{dila21}) if $D\neq 2(d+1)$, then $\bar{b}_0=0$ in order to cancel
the lowest order term on the right hand side of this equation (\ref{dila21}). 
Then (\ref{AF1}) implies the trivial solution, 
$\bar{\phi}_0=\bar{c}_0 = \bar{b}_0=0$. So, we must set $ D=2(d+1) $.
Then the first equation of (\ref{AF1}) is written as
$\bar{\phi}_0=d\bar{c}_0$. From this we can see that if Yang-Mills theory
is asymptotic free then the coupling constant of the sigma model is 
asymptotically infinite and vice versa. This is the reflection of the
relation given in (\ref{coupl}).
For $\gamma\neq 0$ case,
only the solution of negative $\bar{\phi}_0$ is possible as shown below.

Namely, we obtain $\bar{b}_0=-1/2$ and $V_c(T_0)=2d\gamma\bar{\phi}_0$ by
considering the next order terms ($O(1/\rho)$) in (\ref{dila21}).
Then $c_0=-1/2$, and this implies the asymptotically infinite sigma
model coupling constant as pointed out above .
The results are summarized as
\beq
  \bar{\phi}_0=-{D-2\over 4} , \quad 
     \bar{c}_0 = \bar{b}_0 = -{1\over 2}, \quad  D=2(d+1), \label{AF2}
\eeq
\beq
     V_c(T_0)=-{D-2\over 2}d\gamma . \label{AF3}
\eeq
As for the tachyon potential, we can see more from the lowest
and the first order terms of Eq.(\ref{tac21}),
which leads to
\beq
  V'(T_0)=0, \quad V''(T_0)=-2d\gamma . \label{AF4}
\eeq
The second equation of (\ref{AF4}) is derived by assuming $\bar{t}_1\neq 0$.
From these, the fixed points given here will be situated at the maximum point 
of the tachyon potential; at point $A(T_0=0)$ or $C(T_0\neq 0)$ in the Fig.1.

As for the value of $T_0$, 
we can check it from (\ref{AF3}) and (\ref{pot}). If $T_0=0$, we obtain 
\beq
  D={10+d\gamma \over 1+d\gamma} . \label{AF5}
\eeq
This is not satisfied for $D=10$ since $\gamma\neq 0$.
Then the fixed point for $D=10$ should be realized at $T_0\neq 0$
which is identified with the point $C$,
or other local maximum with $T_0\neq 0$.
We give one more remark for the case of $D=10$. We obtain 
$\bar{\phi}_0=-2$ and $d=4$ from equations of (\ref{AF2}), 
then we can see that the coupling constant of 4d Yang-Mills theory
decreases at large $\rho$
as $g_{YM}^2\propto 1/\rho^2$. But the expected case is 
$\bar{\phi}_0=-1$. This point might be resolved by considering
an effective coupling constant obtained from the Born-amplitude of
$Q\bar{Q}$ scattering \cite{Min2}. 

Next, we comment on the solutions for $D<10$. It is easy to find the solution
of $\bar{\phi}_0=-1$ at $D=6$. But we find $d=2$ in this case, so
$\bar{\phi}_0=-1$ can not be connected to the favorable behaviour
of the $\beta$-function of the 4d Yang-Mills theory.
For $D=6$ solution, $T_0=0$ is possible 
and all parameters are determined as
\beq
 \lambda_0=8\sqrt{2},\quad \gamma=1/2, 
        \quad e^{-2C_0}=1/4.
      \label{AF6}
\eeq
Then this fixed point can be taken at the point $A$ in the Fig.1,
and the parameters would be determined definitely.
The above solutions are summarized in the Table 2.

\begin{center}
$$\begin{array}{|c|c|c|c|c|c|}\hline
 \mbox{} & \mbox{(D,d)} & \mbox{$\bar{\phi}_0$} & \mbox{$\bar{c}_0$}
                  & \mbox{$T_0$}  & \mbox{n}   \\ \hline
 \mbox{af1}  & \mbox{(10,4)} & \mbox{-2} & \mbox{-1/2} 
                  & \mbox{$\neq 0$}  & \mbox{5}   \\ \hline
 \mbox{af2} & \mbox{(6,2)} & \mbox{-1} & \mbox{-1/2}
                  & \mbox{could be 0}   & \mbox{3}  \\ \hline
\end{array}$$
Table 2. Asymptotic free solutions for $\gamma\neq 0$.
\end{center}

Finally we examine the stability of the tachyon fluctuation
from its linearized equation of Eq.(\ref{tac21}). Near
the fixed point, $\rho\to \infty$, the linear term of $t$
is expressed as
\beq
    \ddot{t}+d\gamma\dot{t} = {1\over 2\rho} V_c''(T_0)t  \, .
         \label{taclin}
\eeq
Since the right hand side is negligible small, then this fixed point
is stable against small tachyon fluctuation for the above two 
solutions.
%%%%%%%%%%%%%%%%%%%%%%%%%%%%%%%%%%%%%%%%%%%%%%%%%%%%%%%%%%%%%%%%%%%%%

\vspace{.5cm}
\subsection{Solutions for $\gamma=0$}

Next, we consider the case of $\gamma =0$. 
In this case, the bulk configuration will largely deviate from the
horizon configuration, $AdS_{d+1}\otimes S^n$,
of the original D-branes given in the 
type IIB string theory. Then we now widely extend the duality relation
of bulk gravity and the quantum field theory on the boundary or the branes.
 
The equations are solved in the form given
by (\ref{asyma}) $\sim$ (\ref{asymd}). Here we notice the following points.
In this case also, the term from the D-brane, the term proportional to
$\lambda_0$,
must be suppressed at $\rho\to \infty$ at least to the order of $\rho^{-2}$
to obtain a 
solution with $\gamma=0$ as seen from Eqs.(\ref{metric11}) and 
(\ref{metric21}). This is because of the fact that the left hand sides 
of these two equations are the order of $\rho^{-2}$.
The situation is the same for the other three equations.
Then the terms of the right hand sides, the first term of (\ref{metc}) 
(the curvature term of $S^n$) and
the second potential term of (\ref{dila21}), will also suppressed as
$\rho^{-2}$ or more. Several possibilities are considered 
here in solving the
equations depending on which terms are retained 
or abandoned in the right hand sides
as the one of order of $\rho^{-2}$. Among those solutions, we show here
two interesting cases.

First we consider the case where the terms coming from the D-branes and 
$S^n$-curvature are retained and the term from the tachyon-potential is 
suppressed. Namely, we assume 
\beq
 \bar{b}_0+\bar{\phi}_0-\bar{d}\bar{c}_0=-1, \quad
   \bar{b}_0-\bar{c}_0=-1 , \quad \bar{b}_0<-1 .
\eeq
Then we obtain
\beq
  \bar{\lambda}_0^2=2(D-2)\bar{a}_0^2, \quad
 e^{-2C_0}=({D-2\over 2(\bar{d}-1)})^2\bar{a}_0^2 ,
\eeq
where $\bar{\lambda}_0=\lambda_0 e^{-2\bar{d}C_0}$ and
\beq
 \bar{\phi}_0=(d+1-{D\over 2})\bar{a}_0 , \quad
   \bar{c}_0={\bar{\phi}_0\over \bar{d}-1} , \quad 
    \bar{b}_0=\bar{c}_0-1 .
\eeq
Here $\bar{a}_0$ remains undetermined. Since $\bar{b}_0<-1$, we can see
that $\bar{\phi}_0$ and $\bar{c}_0$ should be negative. This solution
then could provide the desirable
asymptotic free 4d Yang-Mills theory if we set as
$\bar{\phi}_0=-1$ and $d=4$, which is possible here.
In order to proceed the systematic perturbation in solving the equations,
the right hand sides of the equations to be solved
should be expanded by the power series of $1/\rho$.
For the above setting, we obtain the integer value of $\bar{b}_0$ for
$D=7$ as $\bar{b}_0=-(D-5)/(D-6)=-2$. And other parameters are determined
as $\bar{a}_0=-2/3, \bar{c}_0=-1$ and
\beq
 \bar{\lambda}_0=-{2\over 3}\sqrt{10}, \quad e^{-C_0}={5\over 6}.
\eeq
From these we obtain the $\beta$-function defined as
$\beta_{\lambda}=d\lambda/d\rho$ for the 
t'Hooft parameter $\lambda=Ne^{\Phi}$ as 
\beq
 \beta_{\lambda}= -{25\over 24\sqrt{10}}\lambda^2+O(\lambda^3) .
\eeq
This result is the expected one for the 4d Yang-Mills theory, but the
coefficient $\beta_0=-{25\over 24\sqrt{10}}$ might be adjusted by the
factor coming from the relation between $\rho$ and the mass
scale in the 4d Yang-Mills theory. In this sense, the relative ratios
of the coefficients will be quantitatively important. 

As for the sigma model,
its coupling constant is asymptotically infinite and we have no knowledge
to compare with this result. Then we comment only on the relation
of the mass scale ($1/a$) in the 2d sigma model and $\rho$. When we respect
the
relation $\rho=2\ln a/\alpha$, we must determine $\alpha$ from the linealized
equation of tachyon fluctuation $t$ as setting $t=e^{\alpha\rho}$. However,
we obtain $\alpha=0$ due to $\gamma=0$. If we set as
$\alpha=0^{-}$, the large
variation of $\rho$ corresponds to a very small change of $1/a$ and the
coupling constant does not run with respect to $a$.

\vspace{.7cm}

We show the second interesting solution. It is obtained when the terms
proportional to $\lambda_0$ and $V_c$ are retained, and the term of
$S^n$ is neglected. This situation is realized by the relations,
\beq
 \bar{b}_0+\bar{\phi}_0-(D-d-1)\bar{c}_0=-1, \quad
    \bar{b}_0=-1, \quad \bar{b}_0-\bar{c}_0<-1 .
\eeq
In this case, we obtain
\beq
 V_c(T_0)=-{D\over 4}\bar{\lambda}_0^2 , \label{pot5}
\eeq
\beq
 \bar{\phi}_0={\bar{d}\over 2\sqrt{D-1}}\bar{\lambda}_0 , \quad
   \bar{c}_0={1\over 2\sqrt{D-1}}\bar{\lambda}_0=-\bar{a}_0 . \label{sol6}
\eeq
In this solution, both $\bar{\phi}_0$ and $\bar{c}_0$ are positive,
then 2d sigma model is asymptotic free but $\lambda$ is asymptotically
infinite. This solution was expected in the section five as the limit
of $\gamma\to 0$ of the (sol1), where $(D,d)=(10,4)$.
Here several values for $(D,d)$ are possible, but we do not consider
this solution furthermore since the situation of the mass scale for the
sigma model is very different from the case of the linear dilaton vacuum
as mentioned above.
The value of $T_0$ should be nonzero since $V_c(T_0)$ should be
negative and finite as seen from Eq.(\ref{pot5}).
Then the fixed points given here would be situated at point $B$ or $C$
in the Fig.1. The above two solutions are summarized in the Table 3.

\begin{center}
$$\begin{array}{|c|c|c|c|c|c|}\hline
 \mbox{} & \mbox{(D,d)} & \mbox{$\bar{\phi}_0$} & \mbox{$\bar{c}_0$}
                  & \mbox{$T_0$}  & \mbox{n}   \\ \hline
 \mbox{af3}  & \mbox{(7,4)} & \mbox{-1} & \mbox{-1} 
                  & \mbox{could be 0}  & \mbox{2}   \\ \hline
 \mbox{af4} & \mbox{} & \mbox{positive} & \mbox{positive}
                  & \mbox{$\neq 0$}   & \mbox{}  \\ \hline
\end{array}$$
Table 3. Asymptotic free solutions for $\gamma = 0$.
\end{center}

As for the stability of the above two solutions, there is no problem in
the sense that the linearized equation of $t$ does not provide imaginary
solution.
The term proportional to $\lambda_0$, which is representing the D-branes,
was essential to obtain the above two solutions. This point would be important
to obtain a realistic critical behaviour of the Yang-Mills theory,
but the solution of $\bar{\phi}_0=-1$ is restricted to $(D,d)=(7,4)$.

\vspace{.7cm}

We finally comment on the fact that we can get more various solutions
when we take the conditions,
\beq
 \bar{b}_0+\bar{\phi}_0-(D-d-1)\bar{c}_0<-1, \quad
    \bar{b}_0<-1, \quad \bar{b}_0-\bar{c}_0<-1 ,
\eeq
by which we can neglect all the terms
on the right hand sides of the Eqs.(\ref{metric11}) $\sim$ (\ref{tac21})
at least in the lowest order. 
The order of these neglected terms are determined by solving the equations of
lower order, where these terms can be neglected, and the consistency
of the results with the above assumptions are assured. 
But we do not consider this case more since it seems to be unnatural to
determine the lowest order part of the renormalization group equations
without D-branes or $S^n$ curvature.

%%%%%%%%%%%%%%%%%%%%%%%%%%%%%%%%%%

\section{Conclusion}

We have examined the critical behaviours of the Yang-Mills theory and the 
2d sigma model through the non-critical string theory,
which is based on the super-symmetric world-sheet action, in the context
of the holography. 
The Yang-Mills theory 
lives on the D-branes at the boundary of the bulk-space and the sigma model
can be considered on the world-sheet of the non-critical string which could
propagate in the bulk-space. They are coupled each other through the 
bulk gravity. The classical solutions of the gravity with D-branes
provides the renormalization group flows for both theories
simultaneously, and the Liouville field plays the role of
the common mass scale of the renormalization group equations in both theories. 

The fixed points found here can be specified by the
vacuum value of the tachyon ($T$) and the position of its potential.
Since the potential is not known except for the region near $T=0$, 
a schematic tachyon-potential $V(T)$ has been used for the help of
understanding.
In the analysis of the sigma model, the D-branes can be neglected and we
obtain the expected $\beta$-function in the linear dilaton vacuum which
is specified at the minimum of $V(T)$ with $T\neq 0$.
When D-branes are present,
we obtain renormalization group equations
for both theories, 2d sigma model 
and 4d Yang-Mills theory.
The fixed points are obtained by neglecting the scale-parameter
dependence of the running couplings. 
We find that the coupling constants of the two theories are not
independent
and their product is finite at the fixed point. This relation 
implies that
the asymptotic freedom of one theory 
can be seen in the region where the coupling
of the other theory is asymptotically infinite. We could find an explicit 
solution which shows the asymptotic free behavior of 4d Yang-Mills theory 
as expected from the perturbation of the field theory. This solution
is obtained in the limit where the AdS like geometry
disappears. On the other hand,
the coupling constant of the sigma model is
very large, and it can not be "running" when the mass-scale is 
replaced by the physical one of the 2d sigma model.
%%%%%%%%%%%%%%

Many problems are remained to be resolved.
There are many other kinds of fixed points and renormalization group flows
for both the sigma model and Yang-Mills theory. 
They are classified by three schematically typical positions of the 
tachyon potential. Before studying physical properties of these
fixed point, it would be necessary to find a reliable form of the
potential in order to accept these fixed points and the renormalization
group flows as the realistic one.
Secondly, we have neglected the running of the tachyon field for the
simplicity.
If we could find a non-trivial running solution of the tachyon, which 
could connect
two positions corresponding to the different-type fixed points
of the Fig.1, it might be a solution of a renormalization group flow
connecting two
fixed points. This is an open problem here.
Further, we should resolve the usefulness of our analysis at small 'tHooft
coupling region where string-loop corrections are usually expected.

\vspace{.5cm}

\noindent {\bf Acknowledgement}

\end{document}